\documentclass[prl,nofootinbib,twocolumn,showpacs]{revtex4}
\usepackage{graphics}
\usepackage{bm}

\begin{document}
\title{Propagating phase boundaries as sonic horizons}

\author{Tanmay Vachaspati}
\affiliation{CERCA, Physics Department, Case Western Reserve University,
10900 Euclid Avenue, Cleveland, OH 44106-7079, USA.}

\begin{abstract}
If certain conditions are met, a propagating phase boundary can 
be a sonic horizon. Sonic Hawking radiation from such a phase boundary
is expected in the quantum theory. The Hawking temperature for
typical values of system parameters can be as large as $\sim 0.04$ K.
Since the setup does not require the physical transport of material, 
it evades the seemingly insurmountable difficulties of the usual 
proposals to create a sonic horizon in which fluid is required to 
flow at supersonic speeds. 
Issues that are likely to present difficulties that are particular 
to this setup are discussed. Hawking evaporation of the sonic horizon 
is also expected and is predicted to lead to a deceleration of the 
phase boundary.
\end{abstract}


\
                                                                                                 
\maketitle
                                                                                                 
Condensed matter analogs of gravitational systems have brought purely 
theoretical ideas on very large-scale gravitational structures to almost 
within the realm of experimental tests. A key breakthrough was the 
realization by Unruh \cite{Unr81} that black hole event horizons can 
be simulated by sonic event horizons
around ``dumbholes'' in fluid flow. Furthermore, theoretical results 
from the application of quantum field theory in curved spacetimes, 
such as Hawking radiation \cite{Haw74}, should also apply to condensed 
matter analogs, thus opening the door to novel experimental tests of 
gravitational phenomena.

The fluid analog of black holes occurs when sub-sonic fluid flow upstream
changes to supersonic flow downstream. Now the fluid flow downstream is too
rapid to allow sonic perturbations to flow upstream. Sound can travel 
downstream from upstream but not upstream from downstream, resulting in 
a ``sonic horizon'' at the location where the flow velocity equals 
the sound velocity. To the upstream observer, the downstream region
is a hole from which no sound can emanate, that is, it is a dumbhole.
This is the classical picture. As in the black hole case, ``dumbholes 
ain't so dumb'' and quantization of the acoustic perturbations leads to 
sonic Hawking radiation from downstream to upstream. Unruh found that the
sonic Hawking temperature is given by:
\begin{equation}
T_{sH} =\left ( \frac{\hbar}{2\pi k_B} \right )
                    \frac{d}{dr}(-v) ~ \biggr |_{hor}
\label{sonicTHUnruh}
\end{equation}
where $\hbar$ is Planck's constant divided by $2\pi$, $k_B$ is Boltzman's 
constant, $r$ is the radial coordinate, $v\equiv |{\bf v}|$, and the flow 
velocity ${\bf v}$ is assumed to be radially inward. The derivative of the 
velocity is evaluated at the location of the sonic horizon determined by 
$v=c_s$ where $c_s$ is the sound velocity. A numerical estimate gives:
\begin{equation}
T_{sH} \sim 3\times 10^{-7} \text{K} 
             \left ( \frac{c_s}{300 \text{m/s}} \right )
              \left ( \frac{1 \text{mm}}{R} \right )
\end{equation}
where the velocity gradient is taken to be $c_s/R$.
The estimate clearly demonstrates the challenge: even when the fluid 
undergoes tremendous acceleration -- velocity change from 0 to $300$ m/s 
within $1$ mm or, equivalently, $\sim 10^7 g$ where $g$ is the terrestrial
acceleration due to gravity -- the sonic Hawking temperature is extremely 
small. 

The immense practical problems in creating a sonic black hole and 
observing Hawking radiation have been described by Unruh \cite{Unr02},
leading to a very grim picture. The crux of the problem is that
the flow of fluids and superfluids suffers from instabilities 
at the tremendous accelerations that are required for creating a 
sonic horizon with a measurable Hawking temperature. If
one chooses to work with a system with low sound velocity, the
instabilities can be evaded, but then the Hawking temperature
is unmeasurably low.  At the moment these difficulties seem to be 
insurmountable, except possibly in Bose-Einstein condensates 
\cite{GarAngCirZol00,GarAngCirZol01}. (Proposals to test quantum 
processes in the presence of cosmological horizons in BEC analogs
also appear promising \cite{FedFis03}.)

%Experiments have also been proposed in which Bose-Einstein Condensates 
%(BECs) are made to flow with varying velocity 
%\cite{GarAngCirZol00,GarAngCirZol01}.  The Hawking temperature can be
%estimated to be on the order of nano Kelvins in these systems. It is 
%claimed that there are no instabilities in the flow \cite{GarAngCirZol01} 
%and that future experiments in BECs would be able to detect Hawking
%radiation. This possibility currently seems to be the only hope for a 
%fluid flow type of experiment.

For the idea to be described in the present paper, a generalization 
of Eq.~(\ref{sonicTHUnruh}) made by Visser \cite{Vis98} will be very 
important. Visser considered the case when the sound velocity itself 
is a function of space and time and obtained:
\begin{equation}
T_{sH} = \left ( \frac{\hbar}{2\pi k_B} \right )
                 \frac{d}{dr}(c_s-v) ~ \biggr |_{hor}
\label{sonicTHVisser}
\end{equation}
This seemingly small step opens up a conceptually new set of possible
dumbhole realizations because now it is not necessary to accelerate
the fluid; instead one can manipulate the function $c_s (t,{\bf x})$. 

Along these lines, the proposal of Jacobson and Volovik \cite{JacVol98} 
(also see \cite{UnrSch03}) employs a domain wall in ${}^3$He-A to 
produce a variation of the ``sound'' velocity \cite{footnote1}
on microscopic scales ($\sim 500$ ${\text\AA}$). Modest velocities of the 
domain wall lead to an enhanced Hawking temperature. The gain is however 
somewhat offset by the rather low sound velocity ($\sim 3$ cm/s). The 
Hawking temperature 
in their setup is $5 \mu$K, whereas reliable thermometry in ${}^3$He currently
only goes down to $100 \mu$K. Furthermore, gradients in the order parameter 
associated with the presence of the domain wall lead to additional sources 
of radiation that are not related to the sonic horizon. It remains to be seen 
if the Hawking radiation can be disentangled from these other effects.

\begin{figure}
\scalebox{0.40}{\includegraphics{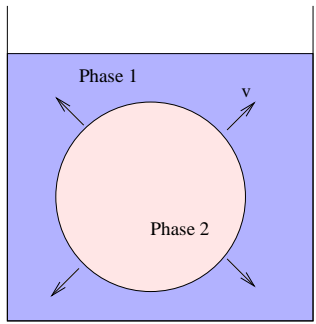}}
\caption{Essentials of the experimental setup described in the text. 
The system is kept at constant temperature and pressure.
}
\label{3dsetup}
\end{figure}

We envisage a setup somewhat related to the Jacobson-Volovik proposal
and pictured in Fig.~\ref{3dsetup}. The container is initially filled 
with a metastable phase (Phase 1) of a material. Suppose that Phase 2
is the stable phase under the conditions of the experiment. Then a
perturbation will trigger bubble nucleation of Phase 2 and this bubble
will grow. The velocity $v$ with which the bubble grows will be subsonic
in Phase 1: $v < c_1$ where $c_1$ is the sound velocity in Phase 1. 
The sound velocity $c_2$ in Phase 2 will be different from that in
Phase 1. We assume $c_2 < c_1$. More crucially we take $ c_2 < v$. 
For the purposes of this paper, we will simply assume that the conditions 
($c_1 > v > c_2$) are met in the system we have chosen. Then sound 
emanating from within Phase 2 cannot catch up with the phase boundary. 
Thus the phase boundary will be a sonic horizon. By the usual 
arguments \cite{Unr81}, sonic Hawking radiation should be seen to 
emanate from the bubble \cite{footnote2}.

The advantage of this setup is that it completely eliminates any instabilities 
associated with supersonic flow. In certain situations, the
bubble wall may push on the material outside. If the system 
is kept at constant pressure, this would cause some expansion of the 
occupied volume. The height occupied by the material would increase by a 
fraction $(\delta V_b)/V$ where $\delta V_b$ is the increase in volume 
occupied by the bubble and $V$ is the volume occupied initially by 
Phase 1. This can be made small by choosing a large container. We can 
also imagine a situation where the transition from Phase 1 to Phase 2,
does not involve a significant change in density. This might happen,
for example, in the case of a fluid to superfluid transition. The bubble 
wall then just marks the location where Phase 1 is being converted to 
Phase 2, without the wall actually exerting any forces on the material 
outside. In this case, there is no transport of material involved. In 
fact, the two phases need not even be fluid, they can be solid, say the
normal and superconducting phases of a Type I superconductor. Our only 
requirement is that we have two different sound velocities and a phase 
boundary velocity that satisfies $c_1 > v > c_2$. If the velocities of 
ordinary (compressional) sound do not satisfy these conditions, one 
could consider some other excitations. However, it is advantageous to 
consider the excitations with the largest velocities that satisfy the 
conditions since then the Hawking temperature is highest.

In this setup, the Hawking temperature can be relatively large because 
the gradient of $c_s -v$ occurs within the phase boundary whose thickness 
will be of microscopic dimensions, given by the inter-molecular spacing, 
or the coherence length of Cooper pairs, or some other such length scale. 
The sound velocity depends on what kind of excitations we are considering. 
We would like to take a large sound velocity, since this gives a larger 
Hawking temperature. With ordinary sound (compressional waves) 
typical sound speeds are $\sim 300$ m/s. This leads to the estimate:
\begin{equation}
T_{sH} \sim 0.04 \text{K} 
           \left ( \frac{\delta c_s}{300 \text{m/s}} \right )
                  \left ( \frac{100 {\text\AA}}{\xi} \right )
\end{equation}
where $\delta c_s$ is the change in the sound velocity across the
phase boundary and $\xi$ denotes the thickness of the phase boundary.
This is quite large compared to Hawking temperature estimates in 
earlier proposals. The thermal frequency is $\nu = k_B T/h \sim 1$ GHz 
and the power emitted is $\sigma_s T_{sH}^4 \sim 10^4$ pW/cm$^2$ 
where the sonic Stefan-Boltzmann constant 
$\sigma_s =  \pi^2 k_B^4/60 c_1^2 \hbar^3$ involves the
sound speed and two polarizations have been assumed \cite{Unr02}.

Now let us consider some specific systems and variations on the
general scheme described above. The first point to note is that
we must necessarily work at low temperature since the Hawking
temperature is of order $0.04$ K. All substances except
the isotopes of Helium, solidify at such a temperature. So we 
need either to work with a solid, or with Helium. If we choose 
$^4$He, we will be working with a superfluid. $^3$He however need 
not be in the superfluid phase. 

\begin{figure}
\scalebox{0.40}{\includegraphics{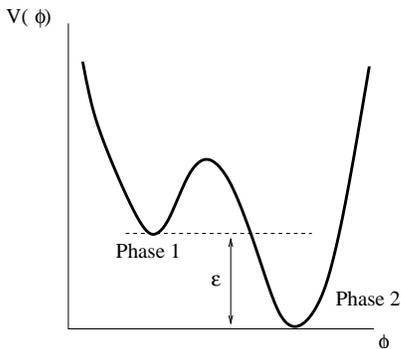}}
\vskip 0.1 in
\caption{A sketch of the free energy $V$ of the order parameter $\phi$.
The force driving the bubble expansion comes from the difference
in energy $\epsilon$ between the two local minima. The drag on
the bubble arises due to scattering of the excitations surrounding 
the bubble with the bubble wall.
}
\label{metastable_pot}
\end{figure}

An existence proof that the conditions $c_1 > v > c_2$ can be met
can be found by considering the Abelian Higgs model in a cosmological
setting. (The non-relativistic version of the Abelian Higgs model 
is the Ginzburg-Landau model used to describe simple superconductors.) 
For a certain set of parameters, the Abelian Higgs model is known to
undergo a first order phase transition, corresponding to Type I
superconductors. Imagine that the system is supercooled, so that 
it is stuck in the metastable state labeled Phase 1 in 
Fig. \ref{metastable_pot}.  Excitations in this metastable state are 
massless gauge particles (photons) and massive scalars. 
If these excitations are in equilibrium, they form a relativistic
fluid for which the equation of state is $p=\rho /3$ (in units where
the speed of light is one). So the sound velocity in the supercooled 
state is $c_1 =1/\sqrt{3}$. If there was no plasma outside 
the bubble, the bubble wall velocity would be the speed of light. 
However, the plasma exerts a drag on the bubble wall, and reduces its 
speed somewhat, and one has $v \lesssim c_1$. (In the similar case of 
the cosmological electroweak phase transition, calculations yield 
$v \sim 0.1$ \cite{Moo00}.) Inside the bubble, in Phase 2, 
all the excitations are massive and non-relativistic. Now the equation 
of state is $p \sim 0$ and the speed of sound $c_2$ is non-relativistic, 
say 300 m/s which is much less than $v$. Therefore the conditions 
$c_1 > v > c_2$ are met, and the bubble should emit sonic Hawking 
radiation. Also note that any heat released by the phase transition
within the bubble does not propagate to the outside since the bubble
wall is moving at supersonic speeds. So oscillations of the order 
parameter in the true vacuum are inconsequential. The only danger
arises from excitations living outside the bubble ({\it i.e.} order 
parameter oscillations in the metastable state) that might scatter
off the bubble wall. However, if the system is sufficiently supercooled, 
the density of excitations in the metastable state is expected to be 
very low.

\begin{figure}
\scalebox{0.40}{\includegraphics{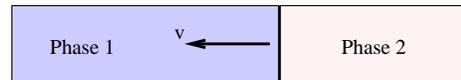}}
\caption{
A one dimensional version of the experimental setup considered in
this paper. The system is kept at constant temperature and pressure.
}
\label{1dsetup}
\end{figure}

A one dimensional version of our setup consists of a cylindrical 
container filled with Phase 1 (Fig.~\ref{1dsetup}). 
A perturbation applied on the right-hand 
end of the cylinder will trigger bubble nucleation of Phase 2. The bubble 
of Phase 2 will start growing and the phase boundary (or bubble wall) will 
propagate to the left. Unlike the bubble of Fig.~\ref{3dsetup}, the area
of the sonic horizon does not change with time and hence the setup
is easier to analyze. 
%In the case of the three dimensional bubble, the
%temperature stays fixed since it is given by Eq.~(\ref{sonicTHVisser}), 
%but the total luminosity will increase with increasing horizon area. 
The disadvantage of the one dimensional setup is that there is more 
interaction of the system with the walls of the container and hence 
there is greater danger of instabilities. In particular, heat released
on the Phase 2 side can be transmitted to the Phase 1 side through the
walls of the container. 

A one dimensional condensed matter situation where a phase boundary 
propagates at supersonic speeds occurs in $^3$He. The boundary of a 
bubble of A phase inside a metastable B phase has been experimentally 
seen \cite{BucSwiWhe86} to propagate at speeds of up to 67 cm/s, while 
fermionic quasiparticles excitations in $^3$He-A propagate at only 3 cm/s.  
On the $^3$He-B side, these excitations are can propagate at 55 m/s and so
the setting seems right for the creation of a sonic horizon. However, note 
that the bubble wall velocity is in the wrong direction in this case -- we 
would like the wall of an A phase bubble, where the sound velocity is small, 
to propagate at high velocity into the B phase where the sound velocity is 
high. It remains to be seen if the A-B phase boundary can provide
measurable Hawking radiation \cite{JacVol98} (also see Sec. 29.3 of 
\cite{Volbook}). 

A novel feature of our proposal is that it is an example of a
``non-driven'' sonic horizon. In other examples, one has to arrange 
external forces to produce a sonic horizon. For example, in fluid 
flow, the fluid has to be driven under pressure. If the external 
conditions are designed to maintain a sonic horizon, the Hawking 
radiation backreaction on the sonic horizon will be lost. In the 
bubble nucleation proposal, however, the procedure is to nucleate 
a bubble of Phase 2 and then leave the system to its own devices. 
If the propagating phase boundary does emit Hawking radiation, it 
will experience a backreaction that will slow it down. Eventually 
the bubble wall velocity will become subsonic with respect to the 
sound velocity in Phase 2. In other words, the prediction is that 
the condition $c_1 > v > c_2$ is quantum mechanically unstable and 
eventually the system will go to a state with $c_1 > c_2 > v$. This 
is no different from the evaporation of black holes due to Hawking 
radiation. 

To estimate the evaporation time, note that the power in the emitted
Hawking radiation is: $\sim \sigma_s T_{sH}^4 A$ where $A = 4\pi R^2$ 
is the area of the horizon when the bubble radius is $R = vt$. The 
energy lost as a function of time is therefore: 
$E_{lost} \sim \sigma_s T_{sH}^4 4\pi v^2 t^3$. The kinetic energy of the 
phase boundary in the absence of the losses would be: 
$E_{kin} \sim \Sigma A v^2$ where $\Sigma$ denotes the surface energy 
density.  Therefore the evaporation time scale is given by equating the 
energy lost to the total kinetic energy:
\begin{equation}
t_{evap} \sim 10^{-6} {f^2}
              \left ( \frac{\Sigma \xi^3 c_1}{\hbar} \right )
              \left ( \frac{\xi}{100 {\text\AA}} \right )
              \left ( \frac{300 \text{m/s}}{c_1} \right ) ~ \text{s}
\end{equation}
where $f \equiv v/c_1$. The evaporation time depends very crucially on
the surface energy density. We would like to have a large surface
density so that the evaporation is relatively slow. The surface
density depends on the system that is being used. However, we can
get an idea for its value by considering the phase boundary to be
a domain wall. For a domain wall in a $\lambda \phi^4$ model, we have:
\begin{equation}
\frac{\Sigma \xi^3 c_1}{\hbar} = O\left ( \frac{1}{\lambda} \right )
\end{equation}
Hence if the coupling constant $\lambda$ is small, the evaporation time 
will be long. This is exactly as for black holes where the Hawking 
evaporation time is long because gravity is so weakly coupled. One way
in which black hole evaporation is different from dumbhole evaporation is 
that the black hole necessarily gets hotter as it gets smaller. The dumbhole 
may get hotter or colder with evaporation, depending on the gradient
of $c_s (t, {\bf x})$ at the location of the sonic horizon $v=c_s (t,{\bf x})$.

If it is difficult to find a system in which the condition $v > c_2$ 
holds, one could consider driving the phase boundary by some external
forces. This might be easy to implement if the substance undergoes a 
phase transition when an electromagnetic field is applied. Then the 
electromagnetic fields could be varied externally to drive the
phase boundary at high velocities. The only danger is that the application 
of dynamical external forces could give rise to additional heating that 
could swamp the signal. We will not consider the ``driven'' setup any 
further here.

We now discuss potential difficulties in implementing the current 
proposal.

The first danger is that the phase boundary may have instabilities
during propagation. Such instabilities need to be avoided since
they will introduce noise in the Hawking emission. Fingering and
other instabilities are system-dependent, and the success of the
proposal will depend on finding a system in which such
instabilities are absent.

An important issue is that, since the phase boundary is propagating, 
eventually the whole container will be filled with the second phase. 
So there is only a limited time $(L/v - L/c_1) \sim L/v$, where $L$ 
is the size of the container, to make measurements. This time had 
better be long enough to detect excitations at the thermal frequency 
$\nu = k_B T_{sH}/h \sim 1$ GHz and we obtain a lower bound on the 
required length of the container: $L > hv/(k_B T_{sH}) \sim 10^{-5}$ cm. 
This condition does not seem difficult to meet in the laboratory where
samples are typically on the scale of centimeters. However, a sample
of 1 cm size will only provide $\sim 10^{-5}$ s to make measurements.

Another potential difficulty is that Phase 1 is metastable and
might decay spontaneously. Hence it has to be protected from
perturbations. However, the Hawking radiation itself will perturb
Phase 1 and the concern is if the system might ``self-destruct''.
To avoid this possibility, Phase 1 should be immune to sonic 
perturbations at the Hawking temperature.

The bubble will have perturbations that are produced during
the nucleation process. Even quantum nucleation leads to fluctuations
on the bubble \cite{VacVil91,GarVil92}. The amplitude of 
fluctuations is suppressed if the bubble surface density, $\Sigma$, 
is large. Hence the effects of bubble perturbations can be reduced, 
as well as the evaporation time made large, by choosing a system with 
a large $\Sigma$.

In summary, we have shown that propagating phase boundaries
can be sonic horizons provided the propagation velocity satisfies
the condition $c_1 > v > c_2$. In systems where this condition is
met, the Hawking temperature can be relatively large, leading to
the exciting possibility that experiments might detect Hawking 
radiation. In addition, in these systems the sonic Hawking radiation 
will lead to evaporation of the sonic horizon, providing a yet
closer analog of gravitational black holes.

\smallskip

\begin{acknowledgments}
I am grateful to Ana Achucarro, Dan Akerib, Craig Copi, Arnie Dahm, Harsh 
Mathur, Rolfe Petschek, Bill Unruh, and Grisha Volovik for discussions and
comments. This work was supported by DOE grant number DEFG0295ER40898 
at Case.
\end{acknowledgments}


\begin{thebibliography}{}

\bibitem{Unr81}
W.G. Unruh,
Phys. Rev. Lett. {\bf 46}, 1351 (1981).

\bibitem{Haw74}
S.W. Hawking, Nature {\bf 248}, 30 (1974);
Comm. Math. Phys. {\bf 43}, 199 (1975). 

\bibitem{Unr02}
W.G. Unruh, in ``Artificial black holes'', 
eds. M. Novello, M. Visser, and G. Volovik, 
World Scientific, River Edge, USA (2002).

\bibitem{GarAngCirZol00}
L.J. Garay, J.R. Anglin, J.I. Cirac and P. Zoller,
Phys. Rev. Lett. {\bf 85}, 4643 (2000).

\bibitem{GarAngCirZol01}
L.J. Garay, J.R. Anglin, J.I. Cirac and P. Zoller,
Phys. Rev. {\bf A63}, 023611 (2001).

\bibitem{FedFis03}
P.O. Fedichev and U.R. Fischer,
Phys. Rev. Lett. {\bf 91}, 240407 (2003).

\bibitem{Vis98}
M. Visser, 
Class. Quant. Grav. {\bf 15}, 1767 (1998).  

\bibitem{JacVol98}
T.A. Jacobson and G.E. Volovik, 
Phys. Rev. {\bf D58}, 064021 (1998).

\bibitem{UnrSch03}
W.G. Unruh and R. Schutzhold
Phys. Rev. {\bf D68}, 024008 (2003). 

\bibitem{footnote1}
The sound velocity that varies within the domain wall is not for 
the usual compressional waves but for certain excitations that 
are very specific to the $^3$He-A order parameter.

\bibitem{footnote2}
To connect with the Unruh setup, it is more convenient to transform 
to the rest frame of the phase boundary. In this frame the fluid is 
flowing with velocity $v$ which is subsonic outside the bubble and 
supersonic inside. Note that the process of cavitation, in which the 
bubble wall simply pushes the fluid and leaves behind an empty cavity, 
does not provide a sonic horizon. In this process, in the rest frame 
of the bubble wall, the fluid is also at rest. 

\bibitem{Moo00}
G.D. Moore,
JHEP {\bf 0003}, 006 (2000).

\bibitem{Volbook}
G.E. Volovik, 
``The Universe in a Helium droplet'',
Oxford University Press (2003).

\bibitem{BucSwiWhe86}
D.S. Buchanan, G.W. Swift and J.C. Wheatley,
Phys. Rev. Lett. {\bf 57}, 341 (1986).

\bibitem{VacVil91}
T. Vachaspati and A. Vilenkin,
Phys. Rev. {\bf D43}, 3846 (1991).

\bibitem{GarVil92}
J. Garriga and A. Vilenkin,
Phys. Rev. {\bf D45}, 3469 (1992).


\end{thebibliography}
\end{document}